# Generative Damage Learning for Concrete Aging Detection using Auto-flight Images

T. Yasuno[a], A. Ishii[a], J. Fujii[a], M. Amakata[a], Y. Takahashi[a]

[a]Research Institute for Infrastructure Paradigm Shift, Yachiyo Engineering, Co. Ltd, Japan
E-mail: {tk-yasuno, akri-ishii, jn-fujii, amakata, yt-takahashi}@yachiyo-eng.co.jp

**Abstract** –
In order to monitor the state of large-scale infrastructures, image acquisition by autonomous flight drones is efficient for stable angle and high-quality images. Supervised learning requires a large dataset consisting of images and annotation labels. It takes a long time to accumulate images, including identifying the damaged regions of interest (ROIs). In recent years, unsupervised deep learning approaches such as generative adversarial networks (GANs) for anomaly detection algorithms have progressed. When a damaged image is a generator input, it tends to reverse from the damaged state to the healthy state generated image. Using the distance of distribution between the real damaged image and the generated reverse aging healthy state fake image, it is possible to detect the concrete damage automatically from unsupervised learning. This paper proposes an anomaly detection method using unpaired image-to-image translation mapping from damaged images to reverse aging fakes that approximates healthy conditions. We apply our method to field studies, and we examine the usefulness of our method for health monitoring of concrete damage.

## 1 Introduction

### 1.1 Related Works

Starting from a climbing robot for inspection in 2000 [1], there has been much research on autonomous robotics for infrastructure inspection [2]; for example, bridge crack detection [3] using unmanned aerial vehicles (UAVs), and so forth. After the deep learning revolution in 2014 [4], vision-based infrastructure inspection techniques have been researched using deep learning algorithms [5]. UAVs as autonomous robotics and vision-based deep learning techniques have been combined for powerful inspection applications [6]-[9].

In the field of infrastructure inspection, there are useful algorithms to detect damages such as object detection tasks and semantic segmentation. However, from a supervised learning standpoint, the damaged class is a rare event and the dataset including such events is always imbalanced, and hence, the number of rare class images is very small. When deteriorations are more progressed, the less events that occurred to enable image collection. Because of this scarcity of damaged data, it is difficult to improve the accuracy of supervised learning in infrastructure inspection. This is one of the hurdles to overcome our underlying problems for infrastructure aging detection for data mining from supervised learning approaches. Instead, this paper proposes an unsupervised deep learning method for aging detection.

Since 2014, the original generative adversarial network (GAN) paper has been cited more than 21,400 times to date (August 14, 2020). Starting from GAN's invention in 2014, the field of GAN has been growing exponentially. Over 500 papers have been published on the topic [10]. GANs may be used for many applications: not just fighting breast cancer or generating human faces, but also 62 other medical GAN applications published until the end of July 2018 [11].

Furthermore, unsupervised deep learning approaches such as the generative adversarial network for anomaly detection algorithms have progressed [12]. When a damaged image is a generator input, it tends to reverse from the damaged state to the health-like state image. Using the distance of distribution between the real damaged image and the generated reverse aging health-like image, it is possible to detect the concrete damage automatically from unsupervised learning.

In the field of infrastructure, there are synthetic augmentation studies to map from the structure edge label to damaged images such as concrete crack and rebar exposure [13]. Here, the conditional GAN framework pix2pix [14] is one of the most successful one that uses paired images, but the image-to-image relationship, i.e., the one-to-one relationship, is a strong constraint for dataset preparation. In particular, we could not collect any unseen damaged image not yet experienced.

In contrast, the CycleGAN framework is flexible for the case of mapping unpaired images from domain-A to domain-B. Even if each number of domain images is different, the CycleGAN framework enables the optimization of end-to-end unsupervised learning. This paper proposes an anomaly detection method using the unpaired image-to-image translation framework CycleGAN, mapping from damaged raw images to reverse aging images as a fake that approximates health conditions. We apply our method to field studies conducted on two dams in Japan.

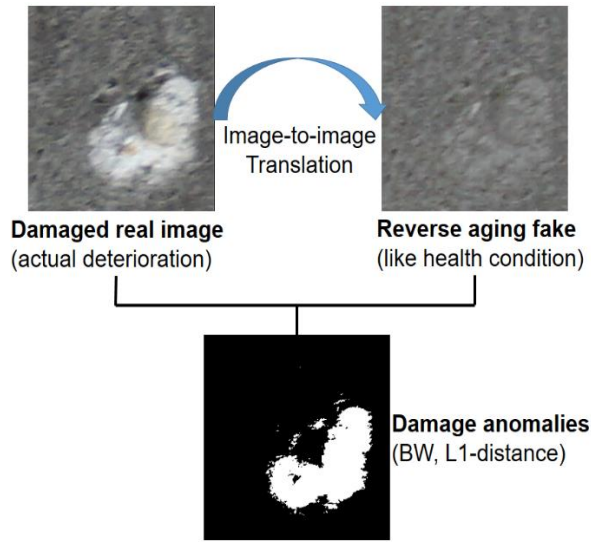

Figure 1. A proposed method using unsupervised generative learning and L1-distance detection.

### 1.2 Unsupervised Aging Detection Workflow

Figure 1 shows an overview of our method using unsupervised generative learning and L1-distance anomaly detection. This method consists of three mainly workflow so as to detect aging infrastructures.

1. Prepare unpaired images dataset: extract from drone images into unit size images and divide two subgroups with damaged and healthy conditions.
2. Train generators and discriminators to optimize the objective function of CycleGAN network mapping functions reverse aging (forward) and deterioration (backward), and cycle consistency.
3. The damage anomalies are visualized by adapting the noise threshold to compute the L1-distance between the real damaged image and the fake output that approximates the health condition using the reverse aging generator.

## 2 Generative Damage Learning Method

### 2.1 Unpaired Dataset Preparation

Auto-flight images captured by a drone have 43 images with a pixel size of 6,000 × 3,000 at the upper left side of dam-1 in the Kanto region. In this study, we set the unit size to 256 × 256. Without loss of resolution, we are able to resize the original size to 5,888 × 2,816 because the resized width is 256 multiplied by 23 and the resized height is 256 multiplied by 11, without any remaining surplus. This method of preparation results in 10,879 unit images.

Second, we classified two groups: a damaged group with damaged regions of interest (ROIs) made up of 4,549 unit images, and a health condition group without any damage made up of 6,325 unit images. Furthermore, we classified four groups 1) health condition without damage (3,852), 2) damaged image (279), 3) blurred (353), and 4) repaired region image (69).

Thus, we create an unpaired image dataset based on the minimum number 222 that contains domain-D (damaged) and domain-H (healthy). The domain-D contains pixel elements of both ROIs and background. Similarly, regarding dam-2 in Tohoku region, the auto-flight drone image has a size of 6,000 × 4,000 pixels, from which we extracted 12,600 unit images of size 256 × 256. This leads to another unpaired image dataset of 237 images.

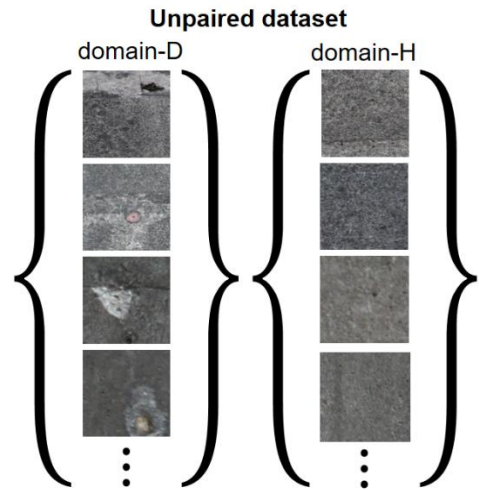

Figure 2. Unpaired images dataset: domain-D (damaged) to domain-H (healthy) translation using CycleGAN.

### 2.2 Damaged-to-Normal Image Translation: Reverse Aging via CycleGAN

Figure 3 shows an overview of our applied CycleGAN framework mapping the "reverse aging" (forward) function R : D → H and the "deteriorated aging" (backward) function F : H → D.

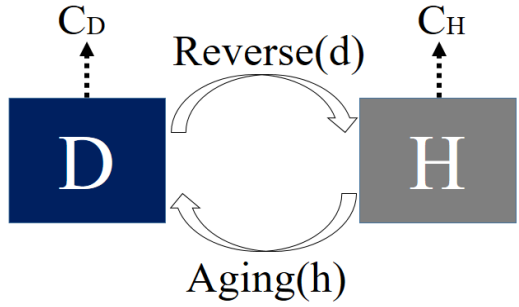

Figure 3. Our applied CycleGAN model mapping functions R : D →H and F : H →D. Discriminator function $C_D$ classifies whether the image is a real damaged image or a generated fake one, and $C_H$ discriminates whether it is a real health condition image or a fake output image.

As Equation (1) shows, for each image $d$ from domain D, the image translation cycle should be able to return $d$ to the original damaged image. This is reverse aging (forward) cycle consistency.

$$d \to R(d) \to A(R(d)) \approx d. \quad (1)$$

Similarly, as Equation (2) shows, for each image $h$ from domain-H, both translation cycles R and A should also satisfy aging (backward) cycle consistency.

$$h \to A(h) \to R(A(h)) \approx h. \quad (2)$$

Equation (3) represents the full objective function that consists of the reverse aging adversarial loss mapping R : D →H and discriminator $C_H$ ; the aging adversarial loss mapping F : H →D and the discriminator $C_D$ , and the cycle consistency loss to prevent learned mappings R and A from contradicting each other as follows:

$$\mathcal{L}(R, A, C_D, C_H) = \mathcal{L}_{GAN}(R, C_H, D, H) \\ + \mathcal{L}_{GAN}(A, C_D, D, H) + \lambda\, \mathcal{L}_{cyc}(R, A), \quad (3)$$

where $\lambda$ controls the relative importance. More detailed numerical representation and network architectures of the generator and discriminator are shown as references [16].

### 2.3 De-noise Anomaly Detection L1-distance

Using the prediction output (health conditions as a fake output) and the input real damaged image, we propose an anomaly detection based on L1-distance. In order to detect damage anomalies as a ROI signal (large difference), another background noise (small difference) is reduced using the background noise threshold $\epsilon$ as a hyper parameter. A larger difference means the deteriorated damage, in contrast a small difference stands for the background noise, such as stain and moss-grown concrete surface. Furthermore, we try to perform blob analysis, such as area open, dilate image, and clear border analysis. This paper computes the next seven steps of image processing as follows:

1. Predict reverse aging output (healthy as a fake prediction) using a trained generator
2. Transform RGB into grayscale of real and fake
3. Centralize the median and set the absolute value
4. Visualize damage anomalies more than the background noise threshold $\epsilon$ , which is the maximum peak vector, exceeds the median
5. Area open to delete four connected elements less than $0.3\epsilon$ (background noise threshold)
6. Dilate image with structural element "octagon"
7. Clear image border where the regions are more blight than neighbors with eight connected

## 3 Applied Results

### 3.1 Dataset of Auto-flight images

Table 1 shows the case study field of two dams in Japan from which we collected images of the concrete surface via an auto-flight drone. Dam-1 took 20 years at an early deterioration stage on the public service timeline. In contrast, Dam-2 is located in one of the snowfall regions of Japan, and consequently, 62 years passed so that several large damages have occurred in appearance.

Table 1. Dam field profile to data collect by Auto-flight

| Profile | Dam-1 | Dam-2 |
|---|---|---|
| Form | Gravity Concrete dam | Arch Concrete dam |
| Height | 156.0 m | 94.5 m |
| Length of Levee | 375.0 m | 215.0 m |
| Service years | 20 years | 62 years |
| Region | Kanto | Tohoku |

### 3.2 Early Damage Study: 20 years

#### 3.2.1 Training Damaged-to-health Image Reverse Aging Translation

Figure 4 shows the training process of discriminator loss using the CycleGAN framework in the Kanto region, where the loss values are transformed into the moving average within an interval of 300 iterations and plots after skipping every 10 iterations. This discriminator classifies whether an image is the real image in the domain-H (health condition) or if it is a predicted fake output. At approximately 15,000 iterations, the discriminator recognizes the generated fake image, but after 40,000 iterations, the fake image often fools the discriminator because the generated image approaches the real health condition. This testing took 17 hours.

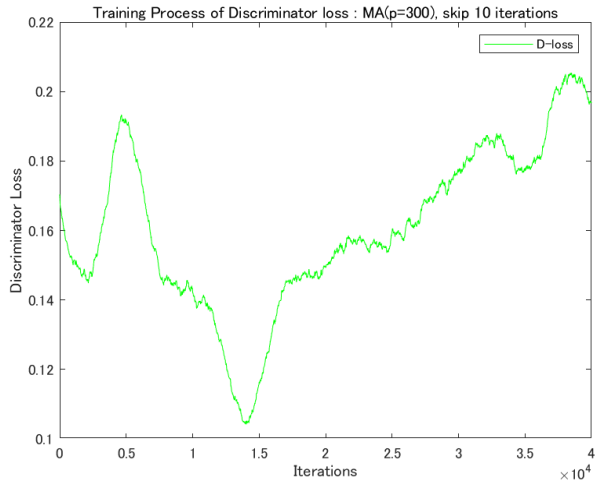

Figure 4. Training process of discriminator loss using CycleGAN framework in the Kanto region.

Figure 5 shows the training process of generator loss using the CycleGAN framework in the Kanto region. This generator transforms from domain-D (damaged) into domain-H (health condition). After 40,000 iterations, the generator approaches a minimum level.

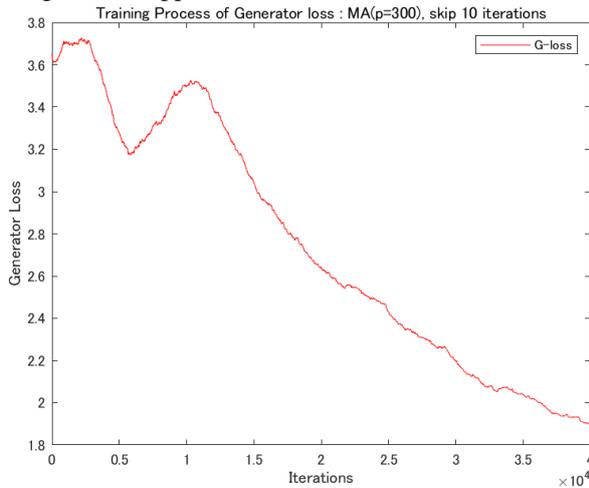

Figure 5. Training process of generator loss using CycleGAN framework in the Kanto region.

### 3.2.2 Anomaly aging detection using L1-distance between raw image and predicted fake

Figures 6 to 9 show the output results of dam-1, such as damaged image (upper-left) and reverse aging "health condition fake" (upper-right) that were translated using trained generator mapping from damaged to health conditions. Both the real damaged and reverse aging fake output are subtracted into a gray-scaled L1-distance mask output (bottom). Although a small noise level remains, our method can detect phenomena such as exfoliation, isolated stone, and sand leak. In particular, Figure 9 shows that sand leak damage is not yet recognized.

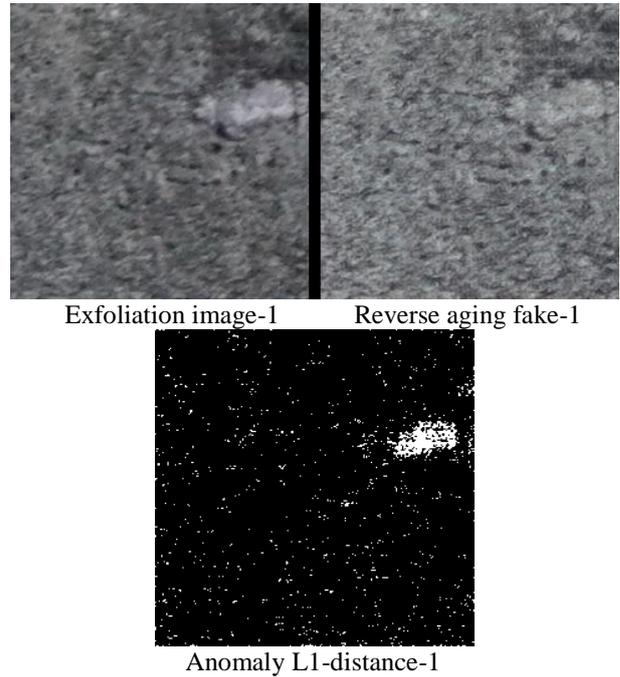

Exfoliation image-1     Reverse aging fake-1

Anomaly L1-distance-1

Figure 6. Exfoliation image (upper-left) and reverse aging "health condition fake" (upper-right) translated using trained generator mapping from damaged to normal image. Grayscale L1-distance mask output (bottom).

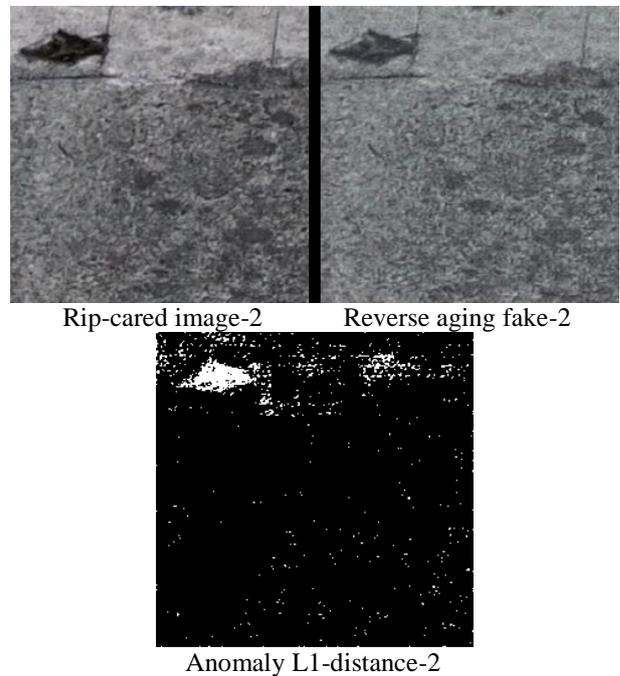

Rip-cared image-2     Reverse aging fake-2

Anomaly L1-distance-2

Figure 7. Rip cared image (upper-left) and reverse aging "health condition fake" (upper-right) translated using trained generator mapping from damaged to normal image. Grayscale L1-distance mask output (bottom).

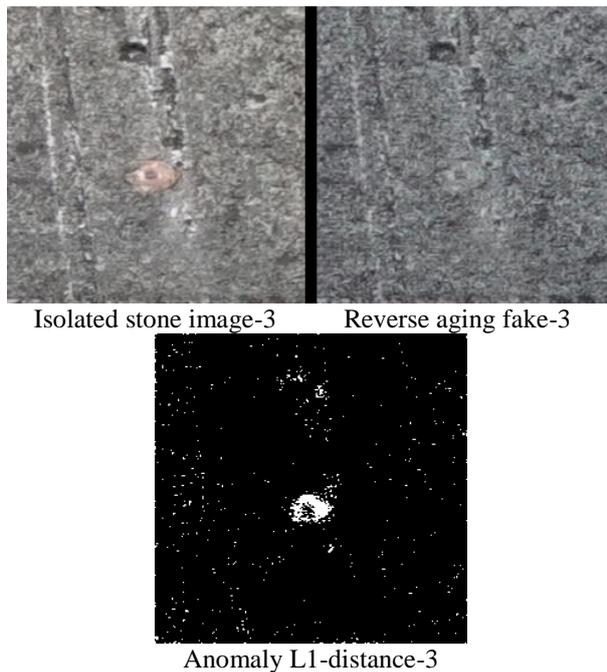

Figure 8. Pre-cause pop-out isolated stone image (upper-left) and reverse aging "health condition fake" (upper-right) translated using trained generator mapping from damaged to normal image. Grayscale L1-distance mask output (bottom).

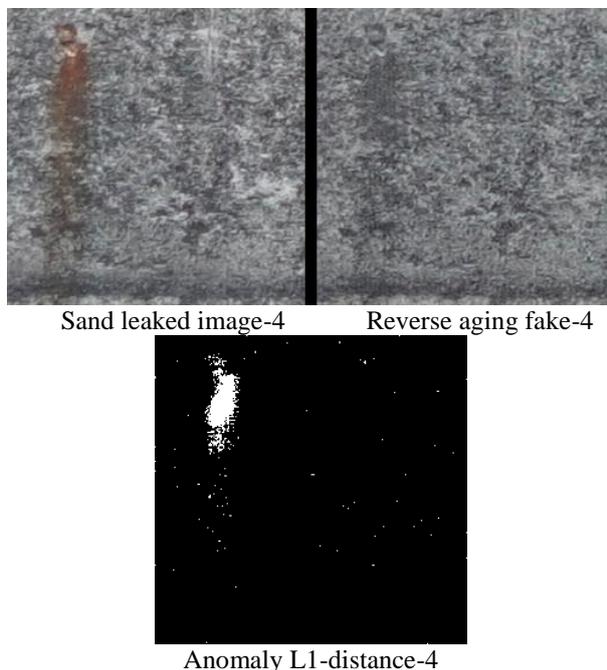

Figure 9. Sand leak image (upper-left) and reverse aging "health condition fake" (upper-right) translated using trained generator mapping from damaged to normal image. Grayscale L1-distance mask output (bottom).

We learn a lesson from this dam-1 case that was 20 years old. Even if the inspection target class is a rare event, by using "reverse aging" generator, any damage anomalies could be detected. In studying a dam over a period of 20 years, our method is useful for "rip care" on the early deterioration stage, such as sand leak and exfoliation, so as to avoid subsequent progressive damage at the prognosis of concrete structures.

### 3.3 Cold Damage Study: 62 years

#### 3.3.1 Training Damaged-to-health Image Reverse Aging Translation

Figure 10 shows the training process of discriminator loss using the CycleGAN framework in the Tohoku region, where the loss values are transformed into the moving average within an interval of 300 iterations and plots after skipping every 10 iterations computing. This discriminator classifies whether it is a real image in the domain-H (health condition) or if it is a predicted fake output image. After 20,000 iterations, the discriminator repeats to fool the output image because the generated image approaches the real health condition. It took 17 hours to complete this training process.

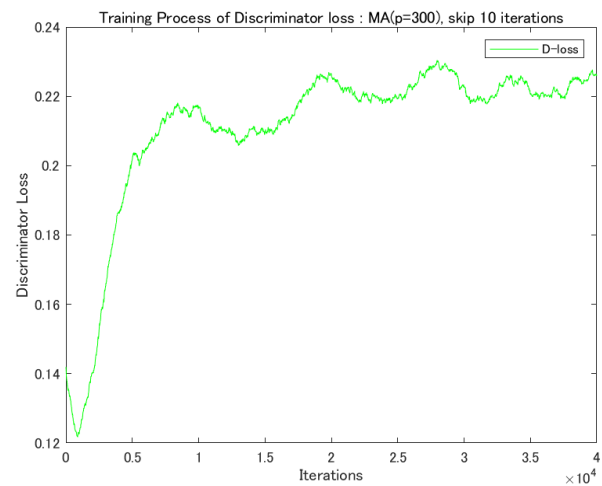

Figure 10. Training process of discriminator loss using CycleGAN framework in the Tohoku region.

Figure 11 shows the training process of generator loss using the CycleGAN framework in the Tohoku region. This generator maps from domain-D (damaged) into domain-H (health condition). After 30,000 iterations, the generator approaches a stable level.

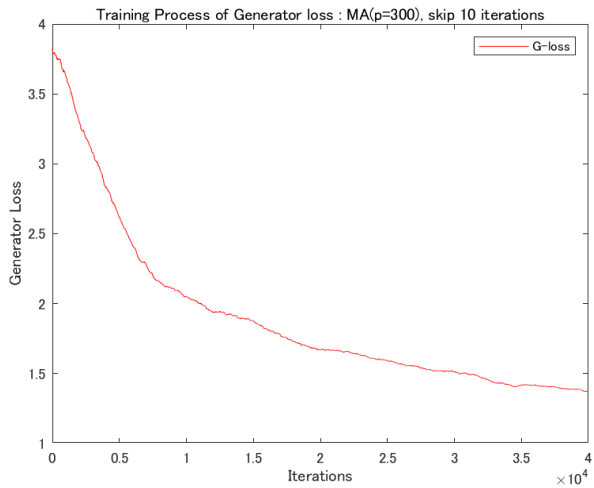

Figure 11. Training process of generator loss using CycleGAN framework in the field of Tohoku region.

### 3.3.2 Anomaly aging detection using L1-distance between raw image and predicted fake

Figures 12 – 15 show the detection studies conducted on dam-2, similar to those on dam-1; the damaged image (upper-left) and reverse aging "healthy fake" output (upper-right) were translated using trained generator mapping from damaged to health condition image. Both the real damaged and healthy fake output are subtracted into a gray-scaled L1-distance mask output (bottom).

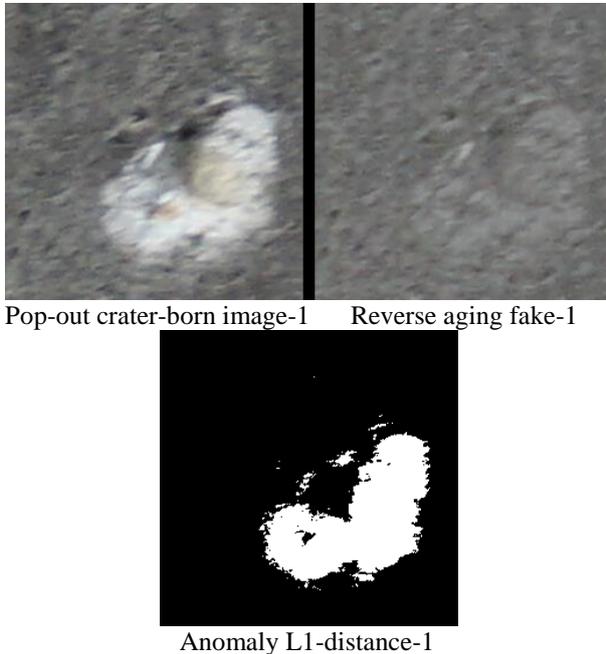

Pop-out crater-born image-1    Reverse aging fake-1

Anomaly L1-distance-1

Figure 12. Pop-out crater born image (upper-left) and reverse aging "health condition fake" (upper-right) translated using trained generator mapping from damaged to normal image. Grayscale L1-distance mask output (bottom).

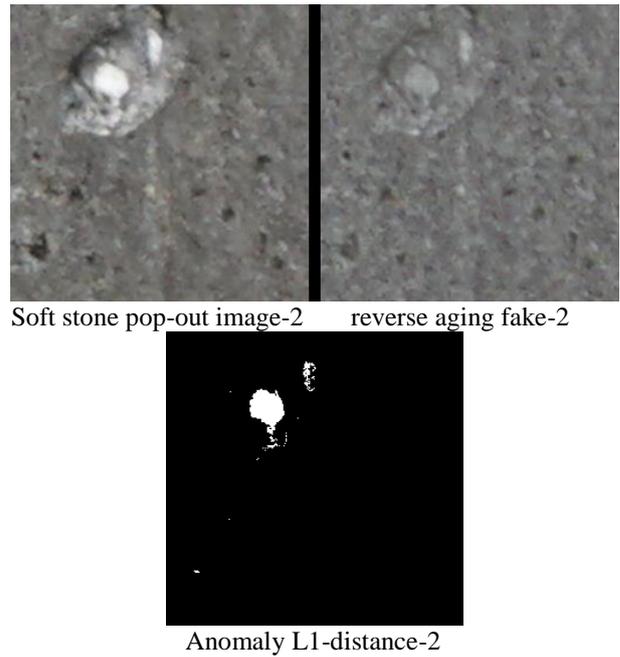

Soft stone pop-out image-2    reverse aging fake-2

Anomaly L1-distance-2

Figure 13. Soft stone pop-out image (upper-left) and reverse aging "health condition fake" (upper-right) translated using trained generator mapping from damaged to normal image. Grayscale L1-distance mask output (bottom).

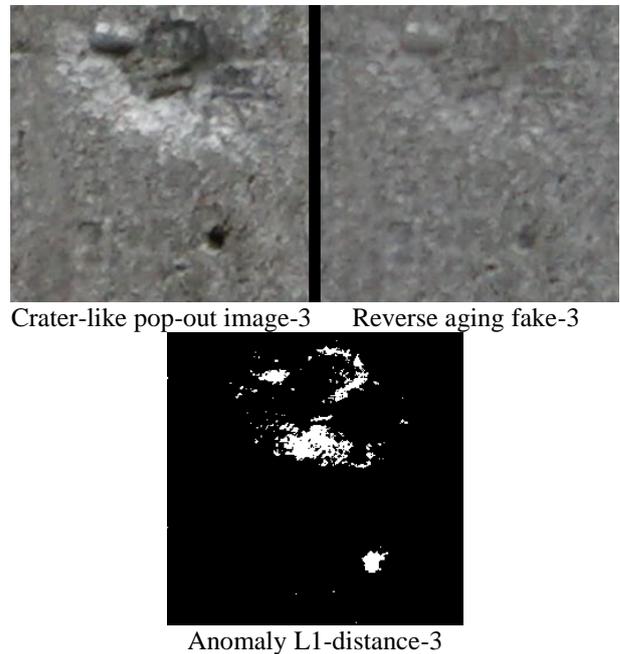

Crater-like pop-out image-3    Reverse aging fake-3

Anomaly L1-distance-3

Figure 14. Crater-shape pop-out image (upper-left) and reverse aging "health condition fake" (upper-right) translated using trained generator mapping from damaged to normal image. Grayscale L1-distance mask output (bottom).

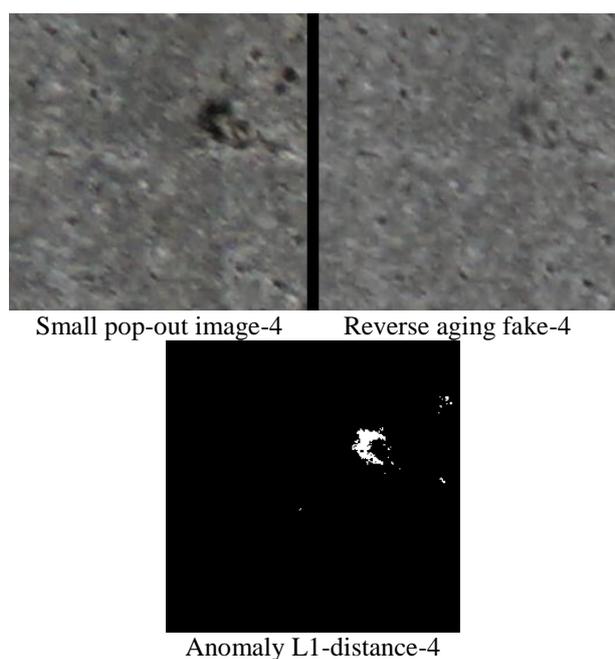

Small pop-out image-4     Reverse aging fake-4

Anomaly L1-distance-4

Figure 15. Small pop-out image (upper-left) and reverse aging "health condition fake" (upper-right) translated using trained generator mapping from damaged to normal image. Grayscale L1-distance mask output (bottom).

## 4 Concluding Remarks

### 4.1 Contributions and Lessons

This paper proposed an anomaly detection method using the unpaired image-to-image translation framework CycleGAN, mapping from damaged raw images to reverse aging predicted fake output as healthy conditions. We apply our method to field studies conducted on two dams, one at its early stages and aged 20 years on the public service, and another one aged 62 years, which is located in a cold region. The CycleGAN framework is flexible because this input only requires an unpaired image dataset such as raw images including a damaged ROI and health raw images without damage. This is a data-mining merit instead of the paired image algorithm, such as that of pix2pix.

In addition to the merit of more flexible dataset, our generative unsupervised approach does not require annotation works supervised for any classes, such as damaged ROIs and background. After we prepare to extract a unit size image and divide a damaged group with another health group, we can use unsupervised deep learning to perform end-to-end training to create an image-to-image translator mapping from the damaged domain to the health (normal) domain. Even if the inspection target class is a rare event, using a "reverse aging" generator, any damage anomalies could be detected. Based on our case studies, in a dam deteriorating over 20 years, our method is useful for "rip care" on the early deterioration stage on the public service of concrete structures, such as sand leak and exfoliation, so as to delay the later progressive damage.

### 4.2 Future Works

It is necessary to optimize a background noise threshold for visualizing the target signal of damage anomalies, and for computing the L1-distance between a damaged image and predicted fake output that approximates healthy condition. This study only considered two case studies of concrete structures. In the future, we will gain further experience with another service periods (e.g. 40 years), much moisture damp locations, steel materials, and another infrastructures such as bridges and tunnels. We will create damage anomaly scores using the L1-distance based on the generator of GAN, in addition to featurematching based on the discriminator. We will also attempt to build another framework that contains a parallel encoder for more efficient damage detection.

**Acknowledgments** We would like to thank Shinich Kuramoto and Takuji Fukumoto for providing practical deep learning MATLAB resources.